# VALUE-CHAIN ORIENTED IDENTIFICATION OF INDICATORS TO ESTABLISH A COMPREHENSIVE PROCESS IMPROVEMENT FRAMEWORK

Ute Riemann[1]

[1]SAP Deutschland AG & Co. KG

## ABSTRACT

- *The process development and optimization potential needs to be driven by the individial coporate value chain.*
- *The identification of this specific value chain and the related indicators is essential to limit the scope of any analysis and optimization to the core business*
- *The process framework consisting of clearly defined value chain, the related processes and the corresponding indicators is a pre-requisite for a meaningful and efficient process analysis and continuous process optimization.*

## KEYWORDS

*Process framework, value chain, business processes, indicators.*

## 1. INTRODUCTION

The need for continuous management and improvement of business processes as a core duty corporate management has to fulfill is a given. However, in real business it is still in discussion what the most useful approach is to establish an efficient business process management for continuous process improvement. The following approach is intended to provide a comprehensive but manageable approach to focus on the management of the processes of highest value for the company and thus creating the most efficient and effective business process management as possible.

The fundamental structure is the value chain as the most comprehensive process within a company. In this approach one key proposition is to rely on the corporate value chain as the fundamental structuring backbone for all business process analysis and improvements. The value chain is the path forward each and every product or service takes from a supplier to the end-customer. The value chain is a company model deployed by Porter focusing on cross-functional orientation in the company. Porter's value chain model is structured by primary activities and support activities.

The analysis of the company and process performance covering the complexity of the value chain requires the understanding of the complexity of the value chain, the related processes and the links of processes and activities within and beyond the value chain. The link between these various activities ensures that the value chain is not a collection of single activities. Moreover, the effectiveness and efficiency of a value chain is expressed by the interaction of these activities within the value chain and to activities belonging to other value chains.

To focus on the core processes and its activities it is a pre-requisite to formalize the identification of the core processes of the value chain to focus on the core processes that are essential for the individual company's business and the root cause for any company success.

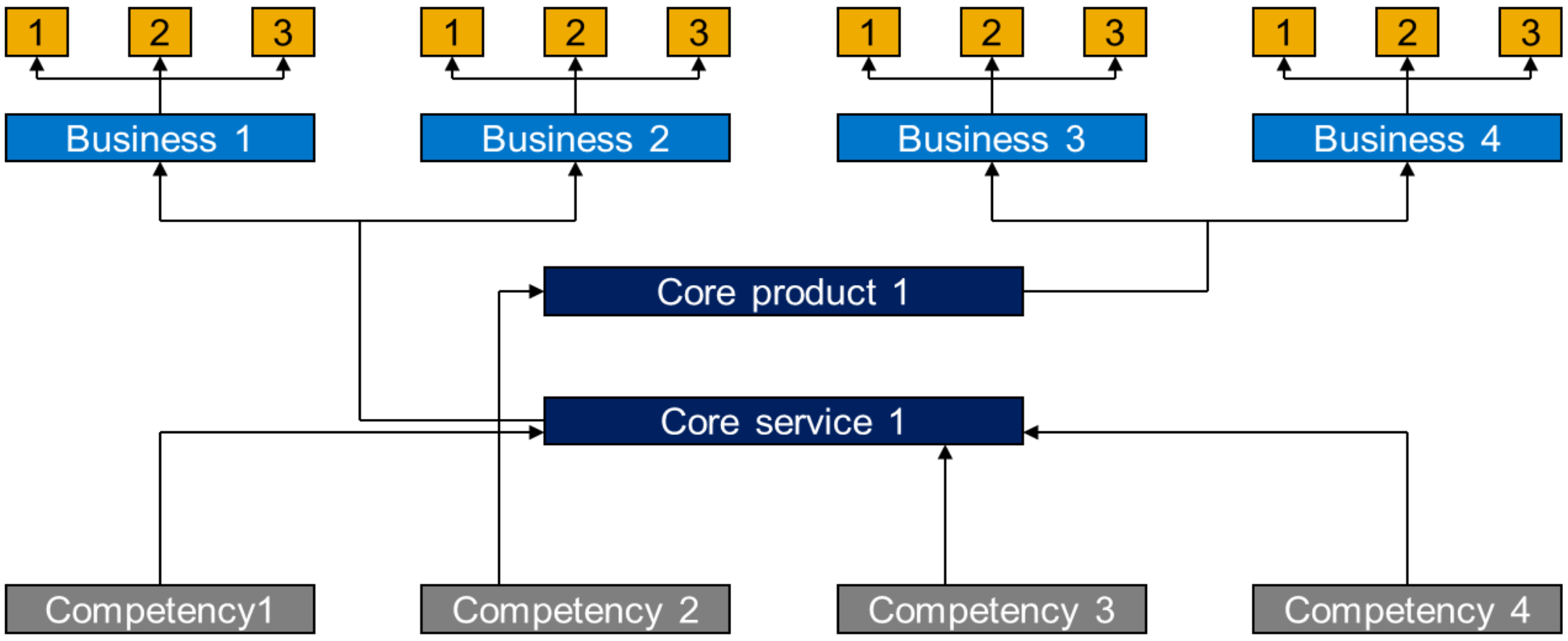

Figure 1. The competencies make the link to the elements of the value chain transparent

With the use of the proposed methodology it becomes relatively easy to allow a value-driven segmentation the value chain with the identification of the most important (= value creating) processes, the corresponding identification of appropriate indicators and measures. Having set this basis we can clearly state the link towards a continuous process improvement. There are several scientific papers available that do consider the value chain for as the key element explaining process improvements. However these papers have not considered the integration of the company's P&L and thus let the corporate strategy and the business impact drive the process improvements. In addition these approaches consider the value-chain as a generic-monolithically basis which is an insufficient view on the value chain to derive the main indicators and measurements from a value chain. The approach stated herein proposed to segment the corporate-specific value chain from a value-based perspective. Differing from Porters process-driven segmentation in core and supporting processes the value-based segmentation allows to focus on those processes and related indicators that provide the majority of value to the company's individual business. However with the value-based segmentation the value chain is individualized and put in a flexible framework considering the processes and its interactions. This does include the links towards KPIs as an expression of the corporate strategy to allow the measurement of the companies (processes) efficiency and effectiveness. Even though key performance indicators (KPIs), both financial and non-financial, are important components of the information to explain a company's progress towards its stated goals, they are sometimes not well used. What makes a performance indicator "key"? And what type of information should be provided for each indicator to measure a company's value chain?

Consequently we need to better integrate the understanding of the KPIs and the link between value chain, P&L and the corporate strategy into the approach which will then be a fundamental differentiator to the existing approaches delivering a significant but dedicated value-add in focusing on the right processes within the value chain and though within the company. This approach allows a unique value-driven segmentation of the corporate value chain as a basis for the identification of the most crucial and important indicators and measures.

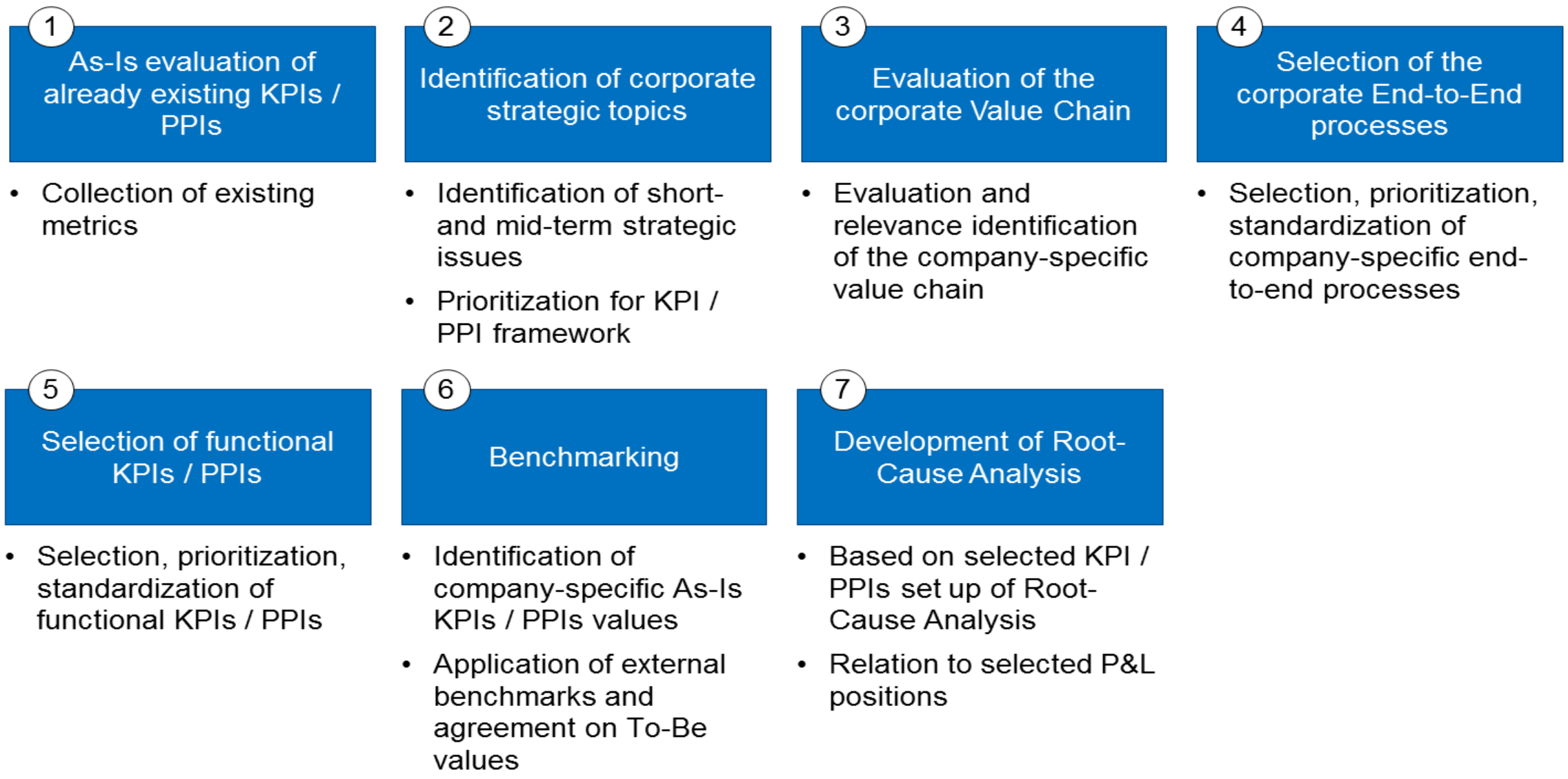

Figure 2: High level approach

The following paper will demonstrate the approach of a value-based segmentation of a company-specific value chain as a basis for the identification of indicators of core relevance. Furthermore I will outline the benefit of this approach in regards to a better and more transparent corporate management based on these indicators and measurements. This approach will be proving with a concrete example of a value chain an on detailed level based on the order-to-cash process.

## 2. THE ROLE OF THE VALUE CHAIN AS THE REPRESENTATION OF THE CORE PROCESSES

To analyze a company from a process (efficiency and effectiveness) point of view it is essential to identify the company-specific value chain of the respective organization. Driven from the corporate strategy, the market developments and customer preferences those processes are of core interest that is essential to deliver the value which is requested by the customers / stakeholders to provide the delivery of products and/or services. Consequently all processes with a significant portion of the overall value generation are of core interest. The nature of the value chain is driven by the industry, the business model and the organization of each company and is therefore individually structured. However the core- and supporting processes and therefore the resulting value chain does follow specific process standards and patters which are close to industry-neutral. Even though the supporting processes are as well important to the overall business generation the core processes will remain in the center of the future approach: the definition of these processes are essential for the corporate business and have an ultimate impact to the operational business excellence and drive the corporate-specific value chain

To further evaluate the KPIs/PPIs it is therefore important to identify these core processes to pre-define the corporate value chain and though keep the scope of the analysis to the core business. This process identification answers the initial question which processes are required to meet the customers / market expectations based on a given business model, the market environment and the corporate strategy. The identification of a meaningful set of indicators will

- help to reduce the overall complexity to a manageable set of data and information to set the right decision in regards to process development and optimization

- enable to focus on the core business within the process environment and bring any further discussions in regards to process optimization to an objective basis. The appropriate set of indicators

- allow the measurability of processes in regards to various business scenarios in regards to

    - the relevance for the corporate business results: how important is a (sub= process for the value chain and how valuable is this process for the corporate result in its current status
    - the process optimization based on the cost intensity: how time and resource consuming is the performance of the process

## 3. AS-IS EVALUATION OF ALREADY EXISTING INDICATORS

Indicators are measurements designed to show how well a business is meeting its goals. Having several indicators for a business is important in order to track its progress. Indicators should be merely the quantitative measurements which reflect critical success criteria as defined by corporate strategy and the core business targets of each company. To allow process indicators acting as a link between strategy, processes and analytical data we need to identify the important indicators which fit to the company’s business expressed by the value-based segmented value chain. Only then we generate the ability to succinctly evaluate the performance of the business process and company. In other words: without the right set and measures of indicators this can be an overwhelming task without ability to share relevant data, misleading information and a non-valuable information basis.

The AS-IS evaluation of the existing indicators will give an initial indication what indicators are devised after a company has established its mission statement, goals and objectives In order to understand how the behavior of the business processes and how the process design affects the business performance we have to first analyze the indicators used in performance judgment and estimation of the process landscape efficiency. This step will not only provide transparency in regards to the indicators the company has currently in use but gives an indication how the company aims to measure the goals as this is an important part of running a business. Consequently the indicators should be developed based on the corporate strategy.

This first step needs to be done carefully as the indicators one company chooses will not likely be the same indicator another one chooses, because it will all depend on the industry and what exactly you are trying to do with the business. These should be viewed long term, and should not change frequently. So this step 1 is not only to identify the indicators landscape and its completeness but is the important starting point to understand the company’s goals and its measurements in-depth- therefore this step 1 is closely linked to the step 2.

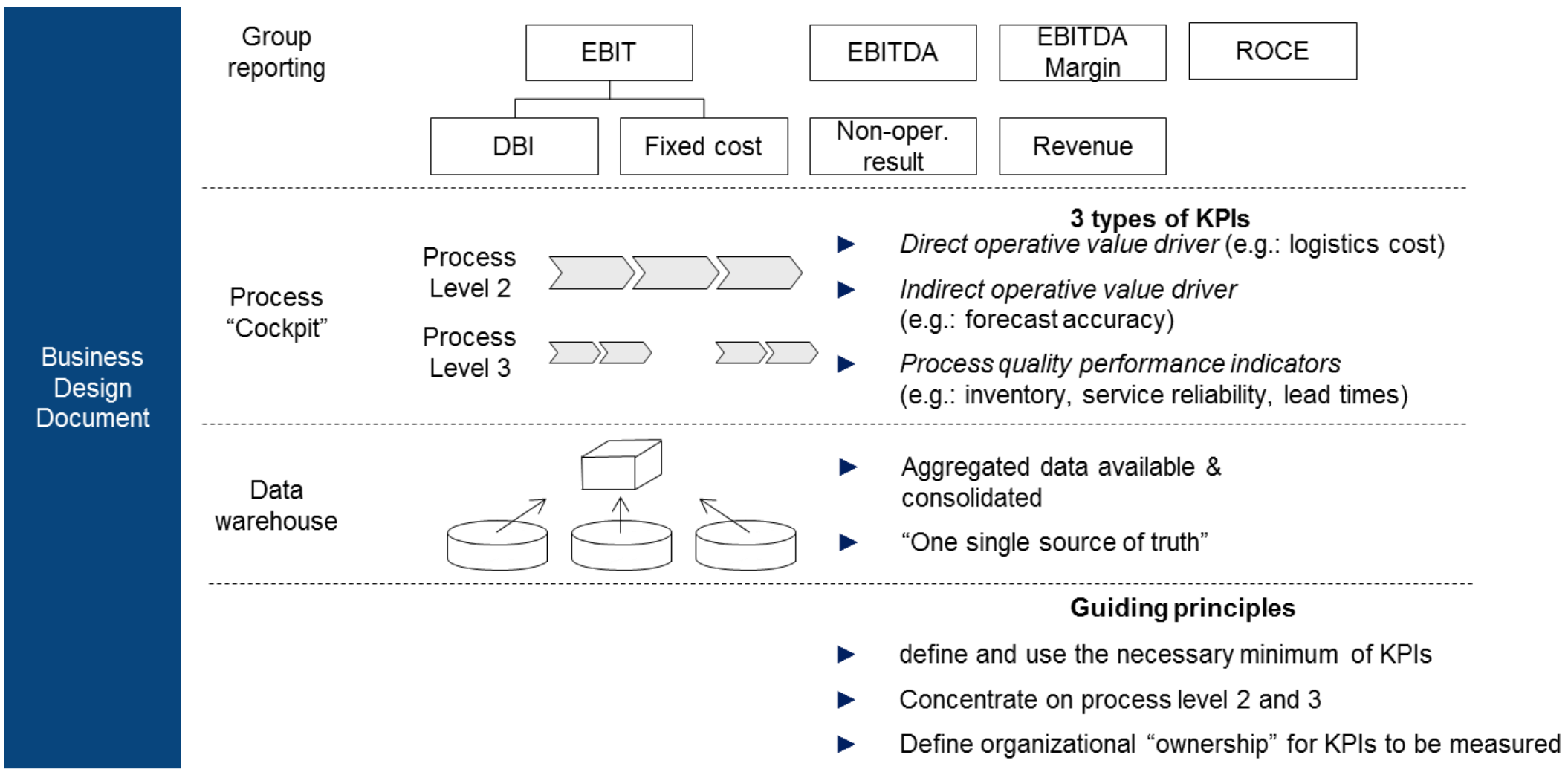

Figure 3: Indicators Tree

Besides the pure understanding of the availability of one indicator it is important to understand the measurement and the link into the overall companies steering model and the targets of the company. The definition of a indicators needs to be stable to allow an accurately measurement of the progress. For example, if you have a goal to increase sales, in the indicator, define how the sales are going to be measured. They could be measured in monetary value, or number of units sold. Factor in product returns and service refunds. Consider whether you are going to deduct them from the month of the sale or the month of the return. It is important to know which metrics you are going to use so you can be consistent throughout and monitor progress more effectively.

## 4. IDENTIFICATION OF COMPANIES STRATEGIC TOPICS

The company's strategic goals are promises to close the gaps between as-is performance and should-be performance. Strategic goals are the gaps that are the most important to close, such that the organisation takes the largest possible steps toward excelling at what it exists to do. The indicators selected shall therefore reflect the goals and those gaps whose are observable how they are developing over time. Therefore it is essential to identify the key goals ad gaps and the corresponding indicators with its benchmarks. The question is how to turn strategic and sometimes lofty goals into measurably meaningful goals. The key to doing that is to explore the possible observable gaps, and then choose the one or two that are absolutely the most important to close. Therefore it is essential to describe the evidence of the goals being attended to focus on the few key goals and of the gap between as-is and should-be closing. Then it is relatively easy to identify the most appropriate indicators and measure them.

## 5. IDENTIFICATION OF A CORPORATE VALUE CHAIN

The evolving customer preferences we need to give those business processes prominence that are essential for the delivery of goods and services to the customers and/or stakeholders. The case is that those processes are of special relevance that to deliver a significant portion of the companies value, that are focused on the fulfillment of the customers' needs and consequently create a perceivable customer benefit. The specification of a company's value chain is different for each industry, business model, organization and strategic goals.
Nevertheless the core and supporting processes and the resulting End-to-End processes and overall value chains have shown process standards and process patterns that are almost industry-independent.

A process is a chain of activities tailored to the provisioning of dedicated services. A process is characterized by a services input, service output, a cycle time, handling time and the use of resources (see Mayer R. (1998), p. 6). The process is an element of a process landscape to gain transparency of the cost driver and the interfaces to other processes.

The starting point for choosing which performance indicators are key to a particular company should be the value chain of each company. To assess continuous business process improvement from a process point of view it is necessary to identify the company-specific value-chain. In 1985 Michael Porter, in his book "Competitive advantage" described the ways in which a company could organize its activities in order to achieve competitive advantage by making it hard for others to copy. His example of a typical value chain included all the organization's external-facing processes, plus their supporting ones. He suggested that once the value chain had been identified, costs could be assigned to the activities to be able to achieve a cost advantage by reducing the cost of individual value chain activities, or by re-configuring the value chain. Based on the corporate strategy, the market development as well as the evolving customer preferences we need to give those business processes prominence that are essential for the delivery of goods and services to the customers and/or stakeholders. The case is that those processes are of special relevance that to deliver a significant portion of the companies value, that are focused on the fulfillment of the customers' needs and consequently create a perceivable customer benefit. The specification of a company's value chain is different for each industry, business model, organization and strategic goals. Nevertheless the core and supporting processes and the resulting End-to-End processes and overall value chains have shown process standards and process patterns that are almost industry-independent. The additional step towards the building of a value chain that goes beyond the classical concept of Porter is to segment the value chain in regards to a value-based value chain driven by process value that will be added to the core business of each company and thus strongly linked to the P&L positions.

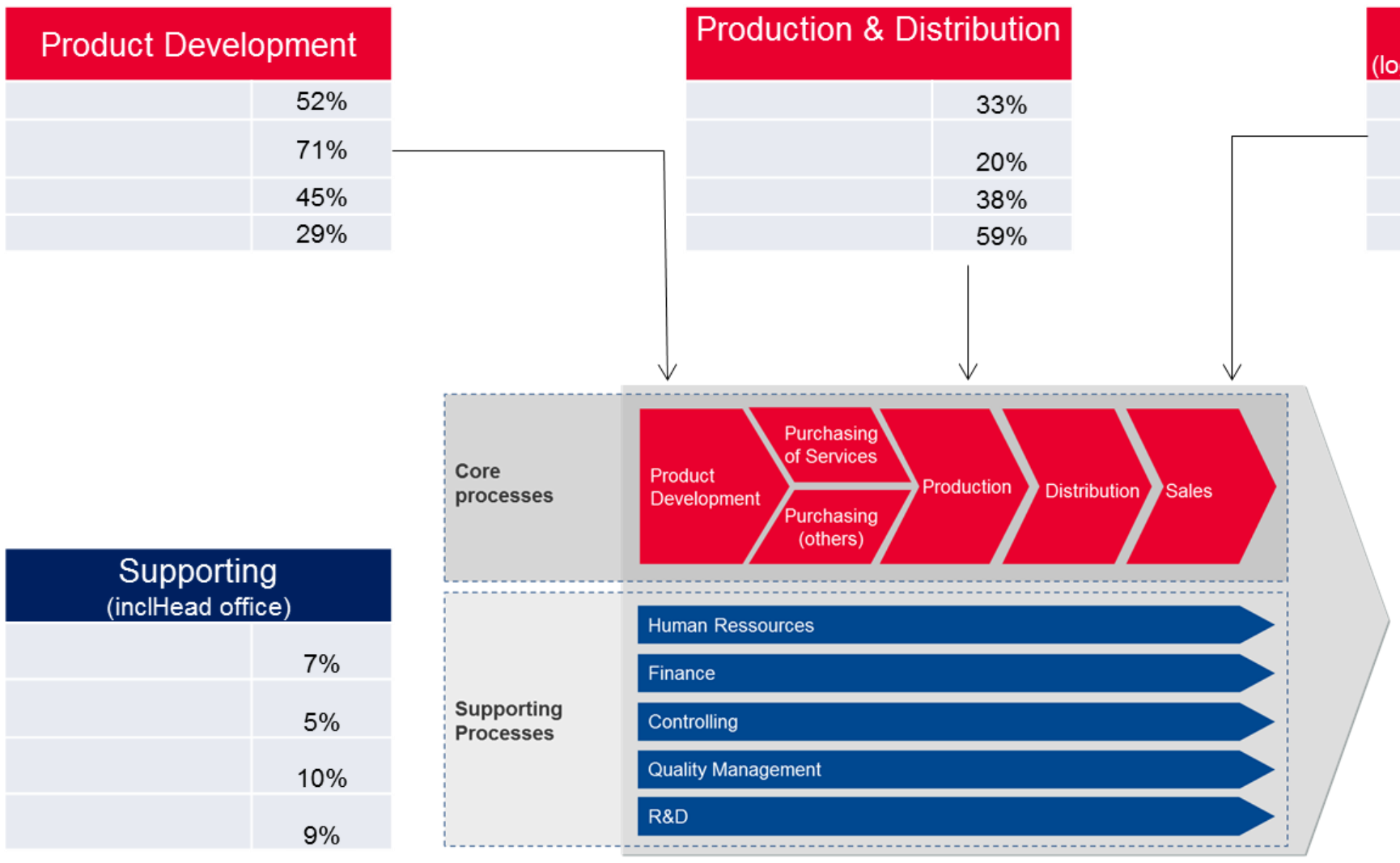

Figure 4: The Cost positions and their proportions within selected companies / segments do give an indicator of the importance and value of the area for business

The cost positions and their proportions within selected companies / segments will help to set the right focus and do give an indicator of the importance and value of the area for the business and thus for the significance of the process and its KPIs. Why bother on this value-based segmentation? As said above, it provides the focused context for all work on an efficient and focused value chain management. But importantly the highest value add is to provide clear definitions, boundaries and dependencies on the most important processes of a company. In a one-step approach the scope is tailored to a manageable but as well most important scope at the same time to let every discussion and activity relate to the core values of a company.

Once this segmentation is achieved the most important End-to-End processes become transparent. These company-specific End-to-End processes are required to formalize the correlation from all operational processes to the (financial) indicators. Even though processes have links and interactions that shall be considered, these processes will serve as the key element to gain transparency of the cost driver and the interfaces to other processes.

A challenge is to identify those indicators that allow assessing processes to assess how the strategies adopted by the company processes and their potential to succeed. In addition, the indicators will to a degree be conditioned by the industry in which a company operates. Indicators presented in isolation from strategies and objectives, or vice versa, cannot fulfill this requirement, and will fail to provide the reader with the level of understanding they need.

In a nutshell the integrated, standardized and harmonized indicator framework will help to steer the integrated, financial and operational analysis and set the basis for basis for continuous challenge and resulting continuous improvement.

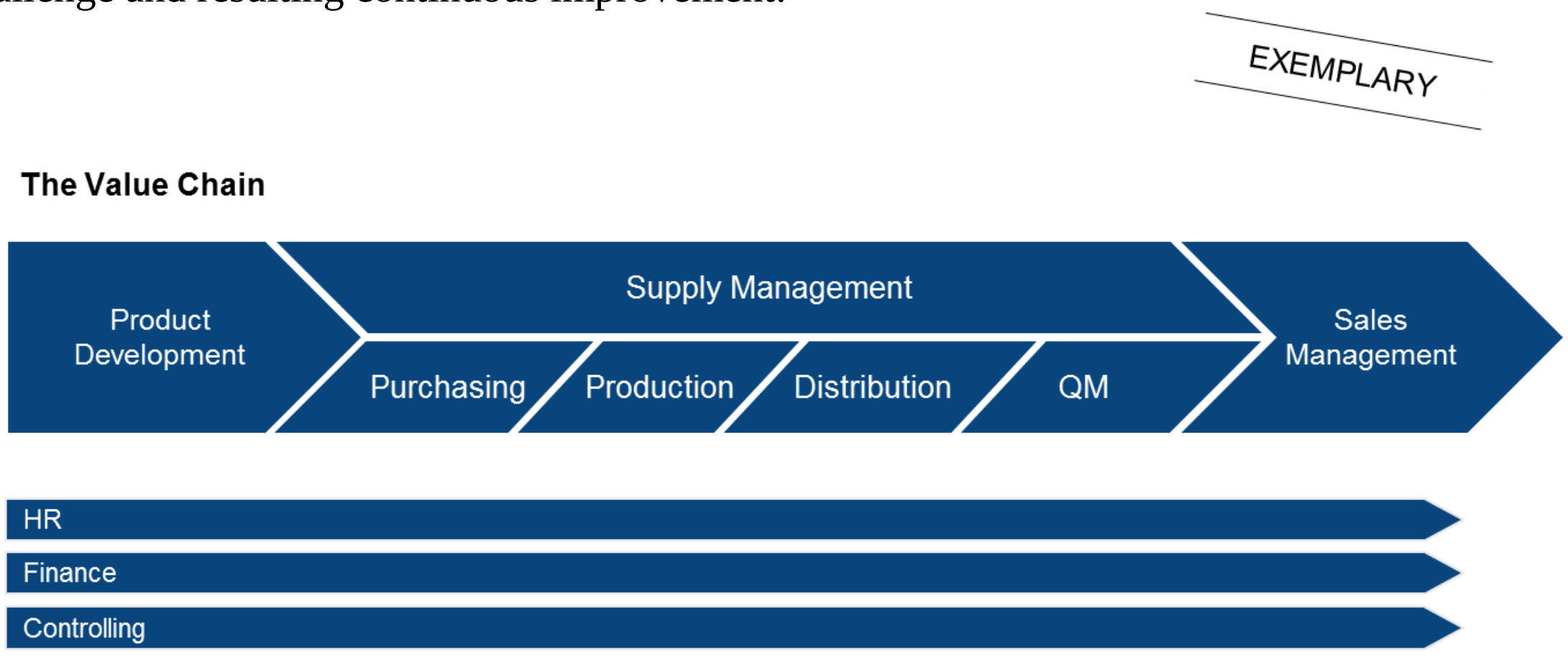

Figure 5: The company-specific Value Chain is decisive to the indicator framework

## 6. SELECTION OF THE END-TO-END PROCESSES

Let me allow starting with a statement: the End-to-End processes are not the same as the value chain. An End-to-End process is a chain of processes that start as the result of a customer trigger and proceeds through until a successful outcome for the customer is achieved. In this article I will therefore use the well-known example of the “Order-to-Cash.” Here a customer contacts a business with the intention of placing an order for a product. This initiates a process that handles the placing of the order, the processing of the order, delivery (possibly manufacture), invoicing, and revenue collection. This is important as the step of the individual value chain shall not be omitted neither the identification of the most important End –to-End processes of a company. So an end-to-end process has some key characteristics:

• It must reflect the customer's view of when they initiate the process and when they get a successful outcome.
• It must reflect the organization's view of when the customer interaction is complete and a business objective has been met.
• It must be capable of being measured, and those measures must take account of the customer view

It is essential to identify those End-to-End processes each company has and then – as a second step to identify those one's that are of ley interest and that are worth measure them specifically. So what are end-to-end processes a company has? As individual the companies and their business are as individual are their End-to-End processes. Even though we might find out that there are typical examples such as Order-to-cash or Procure-to-Pay we might find out the differences when it comes to a detailed process-level analysis. So in any case we need to do the exercise in identifying the company-specific End-to-End processes. The more we think about it, the more we see that the idea of an end-to-end process is rather artificial we need to identify the most important processes with the help of these End-to-End processes – and those ones that gain the most value to the company. This is then the second step: Which of the processes do you focus

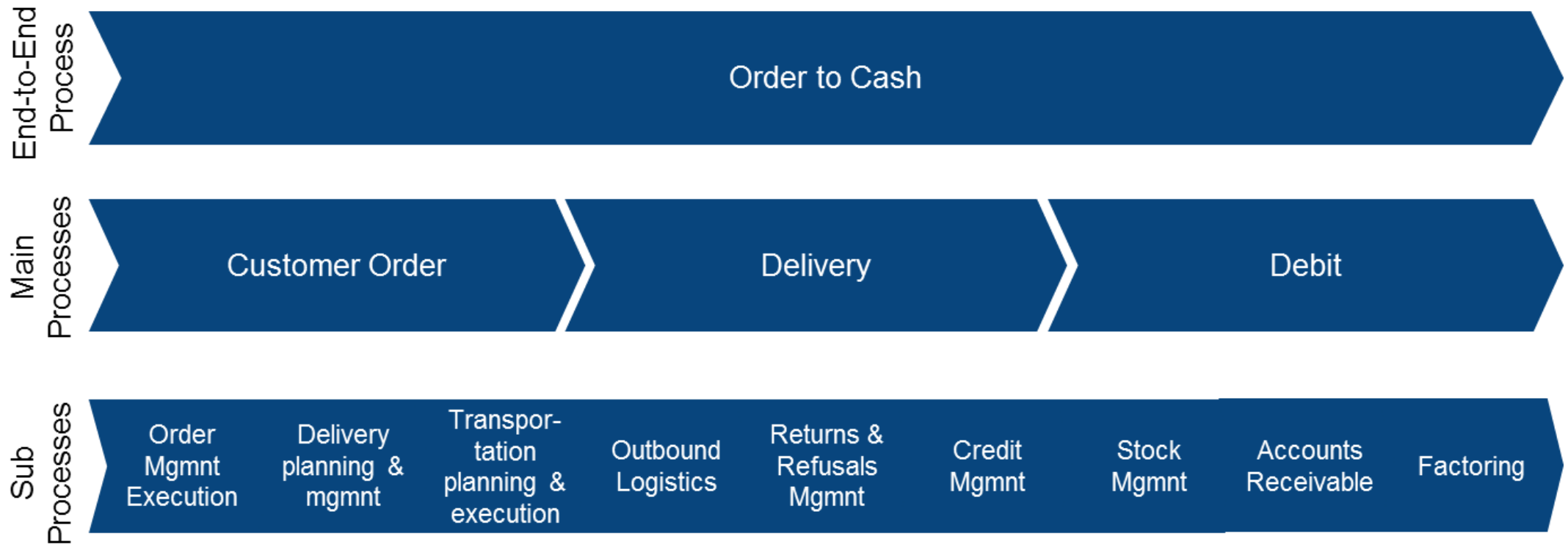

on first? The traditional answer is "Order-to-Cash."

Figure 6: The exemplary view of the Order-to-Cash Process Level 1-3

After all, this is the main process that deals with the customer and generates business revenue, so getting this right is clearly a priority. But if it would be that simple we would not have to investigate the value chain so carefully – wouldn't we? The best "Order-to-Cash" process in the world is of limited value if the business a company is dealing in is focusing on a true consumer market but more on a product-oriented market. Let's assume we are dealing in a monopoly market. What is the impact of any customer-oriented activity to increase sales if there is limited competition? How much benefit you might get in addition if you want to attract a customer that need to buy your products compared to the fact that you might increase you revenue if you focus on the production side. At some point it is obvious, that the company need to drive the importance of an Ed-to-End process from the nature and the core elements of their own business. The key is the P&L that helps to identify the core business activities and though those processes that fit to those P&L positions and is aligned to the business strategy. So there is no key End-to-End process that fits all businesses. The individual corporate business need to be supported by individual End-to-End processes. Even though the processes might sound familiar such as order to cash, procure to pay, or design to deliver in reality, they are specific and support only the business they are designed for. The key step in the use of this approach is the link from the value-segmented value chain to the most important End-to-End processes describing how the most important tasks of a company gets done for a desired objective. An End-to-End process model of could be fairly high level as we are not aiming to drive an execution environment with a much deeper level of detail. The goal should be to capture the few End-to-End process so it can be better understood and interpreted by KPIs. This model –based design goes more toward a complete view of the business interpreted in processes. One part of this is making the complete End-to-End process view more transparent while adding the logic of KPI measurement. To reduce complexity the strong focus on value-driven activities is essential.

The End-to-end process has moreover serve as a starting points for the definition of key indicators I regards to the selection, prioritization, and standardization of End-to-End processes. If achieved the End-to-End indicator tree. The key drivers serve as a basis for the identification of the indicators. Based on End-to-End processes and the respective indicator we are able to analysis the full scope of the business purpose of the process including the process interfaces to cover as well the hand-over across various departments to identify of the optimization potentials.

## 7. SELECTION OF FUNCTIONAL INDICATORS

An effective set of indicators provides feedback in regards to the company's processes and to review the current effectiveness and efficiency. Since these indicators provide assurance on the operational capabilities of a company the defined set of relevant indicators is important to ensure that the right information is gathered. Therefore the real challenge is not only to select those indicators that satisfy efficiency and effectiveness, but also to build the activities needed to meet the levels of process performance required to meet strategic goals. Selecting the right measures is vital for the process efficiency and effectiveness. Even more importantly, the indicators must be built into a performance measurement system that allows individuals and groups to understand how their behaviours and activities are fulfilling the overall corporate goals. The set of relevant indicators will generate organizational behaviours that comply with what is measured which will help to achieve the already defined strategic goals.

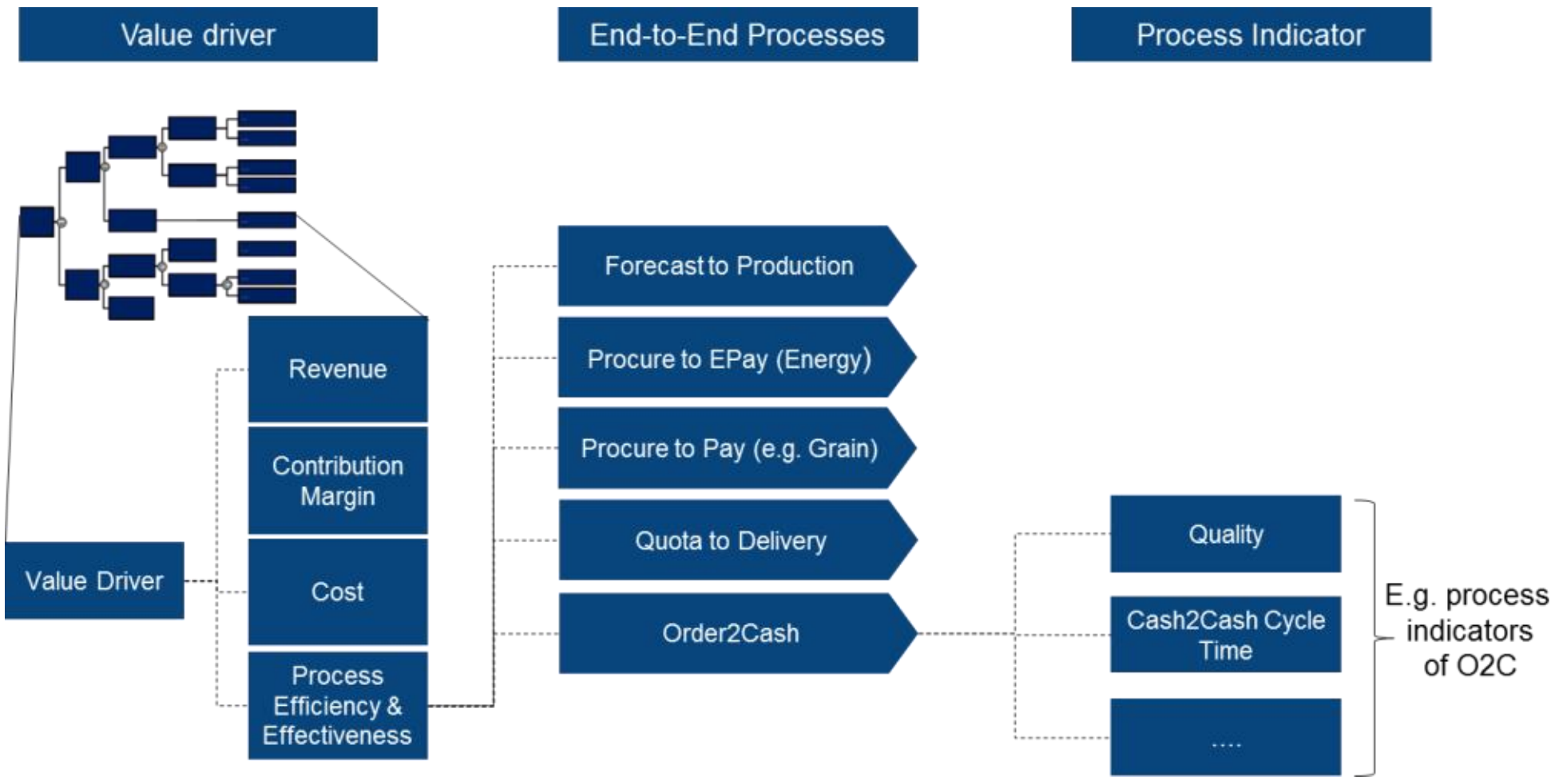

Figure 7: End-to-End indicator logic

Therefore, when defining the indicator the result has to be kept in mind focusing on the outcomes. For the selection of the indicator it is important to overcome the business process segmentation in functional silos and to think in End-to-End process dimensions. This End-to-End process orientation and the definition of the related indicator will help to implement an enhanced way of performance measurement as it treats the comprehensive, customer-facing efficiency in an appropriate way. The key is to build the set of indicator in stages alongside the value-driven value chain. The previously formulated strategic goals create the right perspective for the metrics and provide the overarching framework to define the practical and manageable set of metrics. Here the identified available indicator comes back into the game: what is currently measured? How are these aligned with strategic topics? This lack of connectivity causes dissatisfaction with management reports and criticism of managers who "manage by the numbers." To phrase it in other words: the well-organized set of indicator provide operational measures that have clear cause-and-effect relationships with the core End-to-End processes and thus with the value-driven segmented value chain. Each of these outcomes will build toward the key goals of the company. And these metrics, if well chosen, will be the catalysts for change, providing warning signals to identify ineffective or inefficient processes. The best way to build this link from End-to-End processes to the set of indicator is to map the processes to the identified value chain and all the way up to strategic targets.

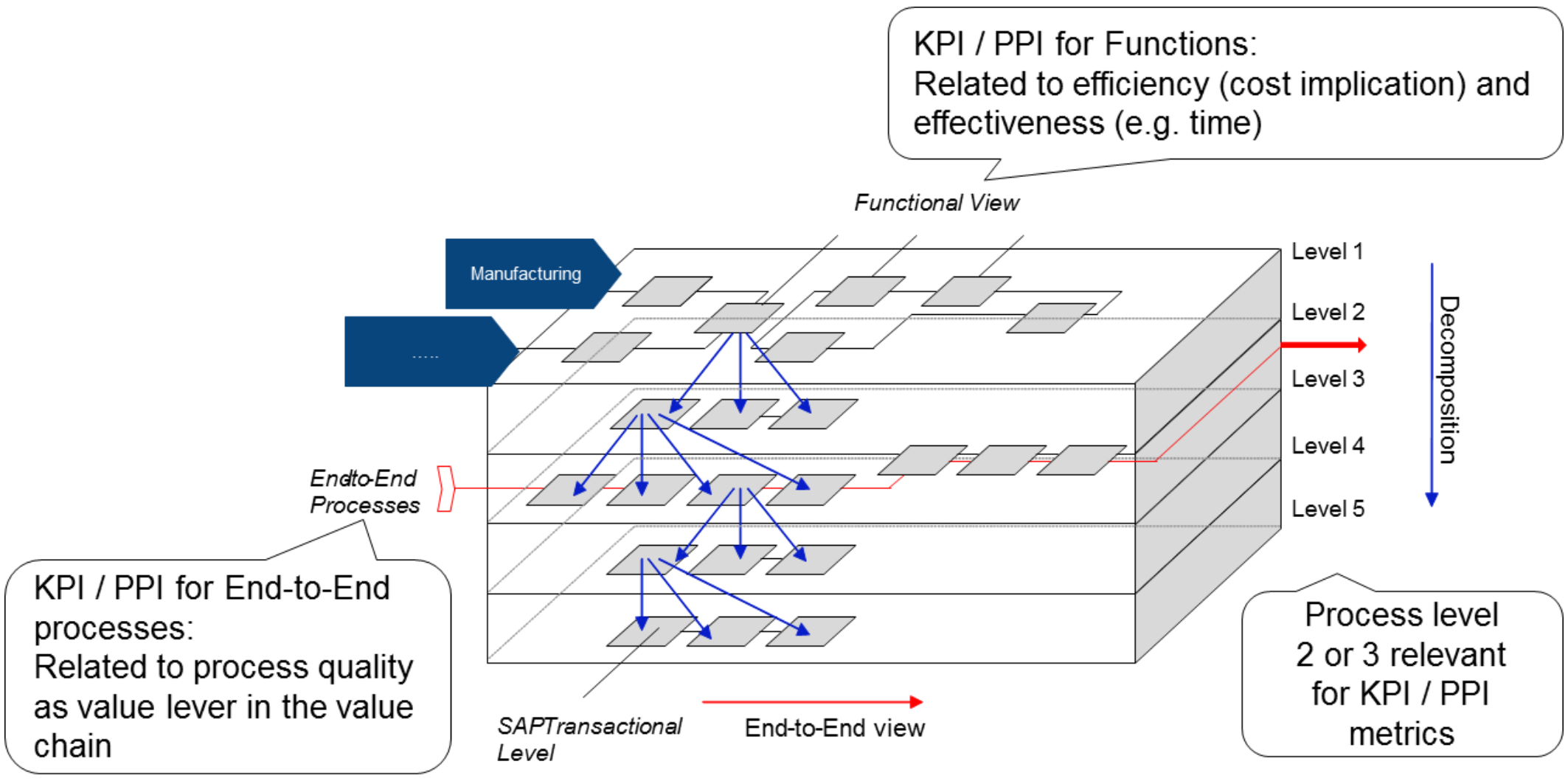

Figure 8: Cross- functional end-to-end processes and functional indicators are measured

As practical application I would like to apply the idea on the Order-to-Cash process. Assuming that the Order-to-Cash process has a significant impact for the company's business the primary objective is to achieve the highest efficiency in managing customer orders with a no-fault-policy. Therefore – from a strategic level- the company initiated a focus on process efficiency and reliability towards the customer to help achieving these targets. The corresponding processes that encompass the Order-to-Cash process have been tailored to these targets. As a result, the required system support has been developed from mainly manual processing to a more or less fully automated processing with top tier performance. However, issues due to production failure still occur randomly and unexpectedly which affects the no-fault-policy. Therefore links to the production processes are affected and needed to be covered within the analysis and by having the appropriate indicators measured.

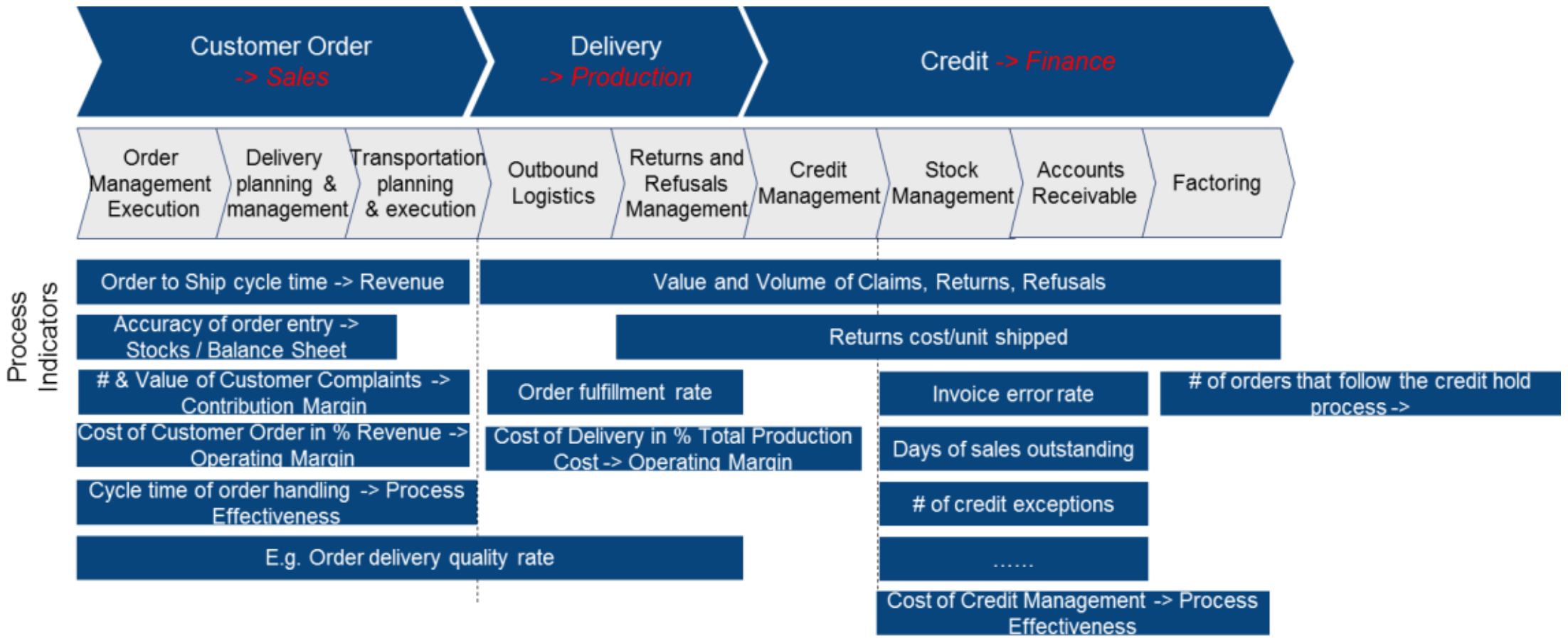

Figure 9: Exemplary dedication of indicators of the Order to Cash Process

This brief example is to outline not only the link from strategic targets to indicator but as well to outline that the focus on key processes is important but that the link to processes that are not of core focus might be relevant as well.

## 8. BENCHMARKING

Benchmarking Performance benchmarked against a relevant external peer group, with an explanation of why these peers were chosen, is considered extremely valuable to users. This provides a clear indication of who management believes the company's competitors to be, as well as setting the company's own performance in the context of a well-defined peer group.

## 9. DEVELOPMENT OF ROOT CAUSE ANALYSIS

Having set the basis with a full transparency in regards to strategic goals, value chain, related End-to-End processes and KPIs the issues a company is facing became obvious. As the intention of this approach is not only to outline issues but to provide an approach to manage issues the step into practice is to solve the problems. Herein I propose to use the technique of a root cause analysis. The root cause analysis is a problem solving process for conducting an investigation into an identified incident or problem. This technique will be applied to our setting as it perfectly allows looking beyond the processes and systems to the immediate problem, understanding the fundamental or underlying cause(s) of the situation and put them right, thereby preventing re-occurrence of the same issue covering processes, procedures, activities, inactivity, behaviors or conditions.

Since we are dealing with a complex situation the root cause analysis required several steps:

- Establish a hierarchy of logically linked information (see figure 10) to make the linkages transparent

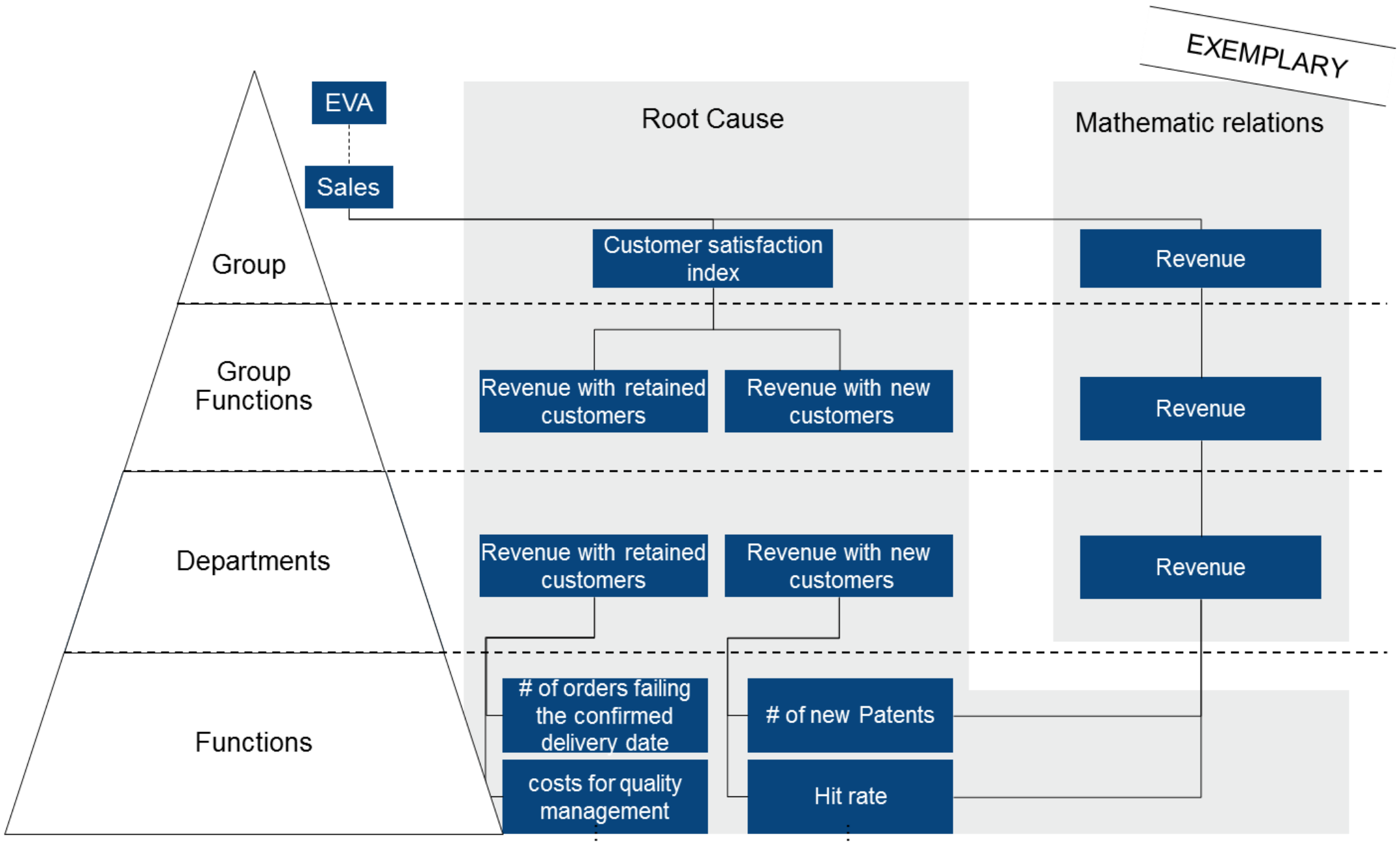

Figure 10: A "Root Cause Analysis" along the hierarchy of selected KPI trees refers to certain P&L positions

- Collation of all available information
- Compile all indicator results or events that do not meet the corporate targets (e.g. based on the benchmarks)
- Investigate possible scenarios & collation of information (for example this may include product or environmental testing or liaison with technical experts and regulatory authorities). List of impacted processes / systems / parties that are affected in regards to the issue
- Draw a conclusion that is the most appropriate for the company and that will led to a roadmap of actions
- Summarize the immediate corrective / long-term corrective as an action plan. Establish an action plan, indicating the action that will be taken to prevent the non-conformity recurring. The action plan should include a defined timescale in which the action will be completed and define who is responsible for completion

It is important that immediate action is taken to correct a deviation from the benchmarking. However, this is separate from the root cause analysis and proposed action plan. The purpose of root cause analysis is to look beyond the immediate non-conformity, to investigate the core processes and thus the corporate value chain. Once this is established, the proposed action plan can focus on ensuring that the system or process can be re-shaped to address the process efficiency and effectiveness in a much better way in the future.

## 10. Conclusion

This paper has shown how a process framework contributes to the continuous process analysis and process optimization. It outlines a path forward to identify the core processes and indicators for each company to allow a stringent and focused process analysis and process optimization. To achieve this, the paper elaborates a value-chain driven approach to identify the related core process with dedicated indicators. It is especially to focus on the core elements of the corporate specific value chain driven from a P&L perspective and at the end-to-end process chains to capture the full value proposition, even if they cross organizational. For the individual valuation of the process steps we have given an example based on a typical OtC. The given example outlines the link from the P&L across various steps to the process identification and the directly related indicators that express the value of each process for the company and emphasizes the core element of the approach.